\documentclass[aps,prl,twocolumn,preprintnumbers]{revtex4-2}
\usepackage{amsmath, mathrsfs, amssymb,amsfonts,amsthm,graphicx, epsf, yfonts, subfigure}
\usepackage[hyperfootnotes=true]{hyperref}
\usepackage{color}
\usepackage{slashed}
\usepackage{setspace}
\usepackage{cancel}
\usepackage{wasysym}
\usepackage{float}
\pdfoutput=1
 \newcommand{\be}{\begin{equation}}
 \newcommand{\ee}{\end{equation}}
 \newcommand{\bea}{\begin{eqnarray}}
 \newcommand{\eea}{\end{eqnarray}}

\newcommand{\beq}{\begin{equation}}
\newcommand{\eeq}{\end{equation}}

\newcommand{\dn}{\mathrm{d}}

\renewcommand*{\thefootnote}{\fnsymbol{footnote}}

\begin{document}


\title{Looking inside a quantum black hole}
\author{Chiara Coviello,$^{1}$ Ansh Gupta,$^{1}$ Robie A. Hennigar,$^{2}$ Kai Shi,$^{1}$ and Andrew Svesko$^{3,4}$}
\affiliation{\vspace{1mm}
$^1$Department of Physics, King’s College London,
 Strand, London, WC2R 2LS, UK\\
$^2$Centre for Particle Theory, Department of Mathematical Sciences, Durham University, Durham
DH1 3LE, UK\\
$^3$Department of Mathematics, King’s College London,
 Strand, London, WC2R 2LS, UK\\
$^4$The Center of Gravity, Niels Bohr Institute, University of Copenhagen, Blegdamsvej 17, DK-2100 Copenhagen Ø, Denmark}

\begin{abstract}\vspace{-2mm}

 \noindent Quantum effects are expected to modify a black hole's interior structure, particularly near the singularity. We show how to extract the scaling exponent of the singularity from the  quasinormal mode (QNM) spectrum of a massless scalar probe in the asymptotic large-overtone limit. We apply our method to a family of exact quantum black holes in (2+1)-dimensional anti-de Sitter (AdS) space and explicitly uncover a transition deep in the interior when quantum effects become dominant.

\end{abstract}

\renewcommand*{\thefootnote}{\arabic{footnote}}
\setcounter{footnote}{0}

\maketitle

\noindent \textbf{Introduction.}
Directly probing a black hole interior is oft assumed to be fatal, as it would require an observer to cross the event horizon. Remarkably, however, the spectra of highly-damped QNMs  provide an indirect means to safely study the near-singularity geometry of black holes \cite{Babb:2011ga,Lan:2022qbb,Konoplya:2022hll,Grozdanov:2026ktq,Grozdanov:2026lnc}. This begs the question whether data about the singularity can be extracted from asymptotic QNMs, and, moreover, if changes to the interior due to quantum effects can be detected, at least in principle. 

Resolving the second part confronts a longstanding challenge. Probing the interior of any black hole requires an exact solution; perturbative corrections to a classical background are insufficient. To wit, suppose  the Schwarzschild metric function has leading corrections like
\be 
f(r) = 1 - \frac{2 M }{r} + \frac{\alpha_1}{r^{n_1}} + \frac{\alpha_2}{r^{n_2}} + \cdots \,, 
\label{eq:blackfactintro}\ee
for $n_2>n_1>1$. Suppose the corrections arise from the same underlying physics, characterized by a common length scale $\ell$. Write $\alpha_i=\lambda_i\ell^{n_i}$ with order-one dimensionless coefficients and assume $M\gg\ell$. At $r\sim\ell$, successive corrections become comparable while remaining small relative to the Schwarzschild term. By the time the first correction competes with $2M/r$, at $r_1=\ell[|\lambda_1|\ell/(2M)]^{1/(n_1-1)}\ll\ell$, higher corrections already dominate it. Thus, without a coefficient hierarchy, truncating the expansion cannot reliably determine the modified near-singularity geometry. Hence, understanding how quantum effects affect asymptotic QNMs requires complete solutions. Attaining said solutions is highly non-trivial, as it means solving the open problem of semiclassical backreaction: how quantum matter influences the classical gravitational field and vice versa.


Even if this issue could be resolved, the next question is what information about the singularity could be recovered. To answer this, it is worth noting that, near the (spacelike) singularity of a generic black hole in $D$-dimensions, before the (semi)classical description breaks down, the geometry has a universal Kasner-form
\beq ds^{2}=-d\tau^{2}+\sum_{i=1}^{D-1}\tau^{2p_{i}(x)}(e^{i})^{2}\;,\label{eq:Kasgeomgen}\eeq
when spatial curvature is negligible. 
Here $\tau$ is an ultralocal proper time coordinate, $e^{i}$ are co-frame one-forms, and the Kasner exponents satisfy $\sum_{i=1}p_{i}=\sum_{i=1}p_{i}^{2}=1$ in vacuum Einstein gravity. The gravitational dynamics features ultra-locality (each spatial point evolves independently of the others) and oscillatory, chaotic dynamics \cite{Belinsky:1970ew}. The full evolution consists of a sequence of Kasner epochs and eras connected
by transitions when the exponents change \cite{Belinski:2017fas}, but satisfy the additive constraints. Quantum effects, when manifested as higher-derivative corrections to general relativity, give rise to new transitions captured by Kasner \emph{eons}, periods of time (longer than epochs and eras) dominated by semiclassical/quantum gravitational physics \cite{Bueno:2024fzg}. Deep in the interior, quantum effects are expected to become dominant, which should correspond to a transition from a classical (Einsteinian) eon to a ``quantum eon''. 

Here we show how to extract the scaling exponent, including Kasner exponents,  of the singularity of a black hole from its asymptotic QNM spectra. For concreteness, we focus on asymptotically AdS$_{3}$ black holes, though our approach easily extends to higher-dimensional spacetimes with flat or de Sitter asymptotics. Applying our method to the \emph{exact} ``quantum'' BTZ (qBTZ) black hole \cite{Emparan:2020znc}, we find the first explicit evidence of a Kasner eon transition between a classical and quantum black hole.

\noindent \textbf{Probing a black hole interior.} For concreteness consider asymptotically AdS$_{3}$ black holes with line element
\beq ds^{2}=-f(r)\dn t^{2}+f^{-1}(r)\dn r^{2}+r^{2}\dn\phi^{2}\;,\label{eq:sphersymmmet}\eeq
though our argument easily extends to higher dimensions. 
We assume at small radius the blackening factor exhibits a power-law singularity
\be  \label{eq:metric_f_ansatz}
f(r) \sim -a r^{-s} \left(1 + c r^p + \cdots  \right) \qquad \text{as } \quad r \to 0 \, , 
\ee
with $s > 0$ and $0 < p < s+ 1$. The curvature singularity at $r=0$ is spacelike for $a > 0$ and timelike for $a < 0$. We probe this background with a minimally coupled massless scalar field $\psi$ with equation of motion $\Box\psi=0$. Separating variables as  $\psi(t,r,\phi) =
    e^{-i\omega t+i m\phi}
    \frac{R(r)}{\sqrt r}$, for azimuthal number $m\in \mathbb{Z}$, the radial equation is brought to a one-dimensional Schr{\" o}dinger-like form
\be 
\frac{{\rm d}^2 R}{{\rm d}z^2} + \left(1 - U(z) \right) R = 0 \, ,  \qquad U(z) \equiv \frac{1}{\omega^2} V\left(\frac{z}{\omega}\right) \, .
\label{eq:schroprob1Dz}\ee
Here we defined $z = \omega \xi$ (assuming $\omega \neq 0$) for tortoise coordinate $\xi(r)\equiv \int_{0}^{r}\frac{\dn r}{f(r)}$, and effective potential 
\begin{equation}
    V(r)
    =
    \frac{f(r)}{4r^2}
    \left[
        4m^2-f(r)+2r f'(r)
    \right].
    \label{eq:brane-potential-general}
\end{equation}
We subject all solutions to the boundary conditions
\beq 
\begin{split}
&R\to 0\;,\quad \;\;\xi\to+\xi_0\;,\\
&R\propto e^{-i\omega \xi}\;,\quad \;\;\xi\to-\infty\;
\end{split}
\label{eq:asympsolns}\eeq
 such that the wave vanishes at the asymptotic AdS boundary ($\xi_{0}\equiv \xi(\infty)$) and is purely ingoing at the horizon.
Away from regions where $U(z)$ exhibits a compensating divergence, as $|\omega| \to \infty$ the potential $U(z) \to 0$, and the large-$|\omega|$ solution is approximated by plane waves.
Nontrivial information about the QNM spectrum is thus localized near the singular points of the potential. 

We will compute the leading and subleading QNMs in the large-overtone limit. Our strategy is to solve (\ref{eq:schroprob1Dz}) as follows. We analytically continue the radial (tortoise) coordinate to the complex plane. Traversing along the Stokes line, $\text{Im}(\omega \xi)=0$, the asymptotic solutions are purely oscillatory such that it is straightforward to impose boundary conditions (\ref{eq:asympsolns}). We solve (\ref{eq:schroprob1Dz}) in a neighborhood near the singular points of the potential and then transport the resulting solutions along a carefully chosen contour in the complex$-\xi$ plane. Matching the solutions near: (i) spatial infinity,  (ii) the origin, and (iii) the horizon, leads to the asymptotic QNM quantization condition. Subleading corrections to the spectrum are obtained by applying perturbation theory to the subleading terms in the potential near one of its singular points. We summarize the main aspects of the analysis, relegating details to the supplementary material.

\noindent \emph{Leading asymptotic modes.} Near spatial infinity, $r\to\infty$, at leading order the master equation (\ref{eq:schroprob1Dz}) reduces to 
\be 
\frac{\dn^{2} R}{\dn z^{2}} + \left(1 - \frac{3}{4 z^2}\right)R = 0 \, ,
\ee
with $z = \omega(\xi_0 - \xi)$.  The general solution is a superposition of Bessel functions. Imposing normalizable boundary conditions and expanding the solution at large argument yields
\be 
R_{(\infty)}(z) \sim A_1 e^{i \left(\frac{3 \pi}{4} - \omega \xi_0 \right)} e^{i \omega \xi} +  A_1 e^{-i \left(\frac{3 \pi}{4} - \omega \xi_0 \right)} e^{-i \omega \xi} \, .
\ee
 Near a non-degenerate horizon $r_{h}$, one has $\xi\sim\log(r-r_{h})$ and the potential vanishes linearly such that 
\be 
R_{(h)}(\xi) \sim B_1 e^{i \omega \xi} + B_2 e^{-i \omega \xi } \, .
\ee
The horizon ingoing boundary condition sets $B_1 = 0$. 
Finally, near the origin the potential at leading order is,
\be 
V_{(0)} = - \frac{(1+2s) a^2}{4} r^{-2(1+s)}\;,
\ee
and the leading-order differential equation takes the form
\be 
\frac{\dn^{2}R}{\dn z^{2}} + \left(1 + \frac{\frac{1}{4}-\nu^2}{z^2} \right) R = 0 \,, \qquad \nu = \frac{s}{2 (s+1)} \, . 
\ee
The general solution is a superposition of Bessel functions, $\sqrt{z} J_{\nu}(z)$ and $\sqrt{z} J_{-\nu}(z)$,
which at large-$z$ has the form
\be
\begin{split}
R_{(0)}(z)  \sim &\left(C_1 e^{-i \alpha_+} + C_2 e^{-i \alpha_-}\right) e^{i \omega \xi} \\
&+ \left(C_1 e^{i \alpha_+} + C_2 e^{i \alpha_-}\right) e^{-i \omega \xi } \,, 
\end{split}
\ee
for $\alpha_\pm = \frac{\pi}{4} \pm \frac{\pi \nu}{2}$. 

\noindent  \emph{QNM quantization.} Inspired by \cite{Motl:2002hd,Motl:2003cd,Cardoso:2004up,Natario:2004jd}, we apply the method of Stokes line matching to attain the asymptotic QNM quantization condition. See \cite{Coviello:2026hdc} and supplemental material for more details. By our contour prescription, we transform the solution near the origin by a $\pi/(s+1)$ rotation in the complex $r$ plane.  Expanding at large argument, matching the solutions on the infinity-to-origin portion of the contour and imposing (\ref{eq:asympsolns}) yields
\be \label{eq:qnm_quant_leading}
\omega \xi_0 = n \pi + \frac{i}{2} \ln \left[2 \cos \frac{\pi s}{2(1+s)} \right]  \, 
\ee
for overtone number $n = 0, 1, 2, \dots $.
Thus, the scaling exponent $s$ of the singularity can be extracted from the offset. Admittedly, our argument relies on an assumption about the global Stokes topology of the complex-$r$ plane; that it be Schwarzschild-AdS-like. We now argue, with the subleading corrections to (\ref{eq:qnm_quant_leading}), this assumption can be relaxed and we can robustly extract the scaling of the metric function near the singularity. 

\noindent \emph{Subleading corrections.} To obtain subleading corrections to the asymptotic QNMs, one treats deviations from the exact Bessel problems that govern the near-singularity and near-boundary problems as perturbations~\cite{Musiri:2003bv, Cardoso:2003vt, Musiri:2005ev}. 
To wit, suppose near a second-order pole the effective potential has a subleading term like
\be 
V(\xi) = \frac{\nu^2 - 1/4}{\xi^2} + b \xi^{q-2} \, .
\ee
The corresponding radial equation (\ref{eq:schroprob1Dz}) for $z=\omega\xi$ is
\be 
\frac{\dn^{2}R}{\dn z^{2}} + \left[1 + \frac{1/4-\nu^2}{z^2}  - b \omega^{-q} z^{q-2} + \cdots \right] R = 0 \, .
\ee
For $q>0$, since $\omega\sim n$ at large overtone, then $\lambda \equiv -b \omega^{-q} \sim n^{-q}$ serves as a parameter to perform perturbation theory. The correction to the QNM spectrum will have the leading behavior $\delta \omega \propto n^{-q}$. This scaling is due to the second-order pole of the potential near the origin and that leading order QNMs go linearly with $n$.

With this in mind, let us identify the subleading corrections to the potential for our metric ansatz~\eqref{eq:metric_f_ansatz} and those arising from the angular momentum $m$ contributions to the potential. Expanding near $r = 0$ and using~\eqref{eq:metric_f_ansatz} we find (see supplemental material for details)
\begin{align}
    V_f &= \frac{1}{\xi^2} \left[- \frac{1+2s}{4(s+1)^2} + \frac{cp(p-s)\rho^{p}}{2 (s+1)^2 (s+p+1)} + \cdots \right],
    \\
    V_m &= - \frac{m^2}{a(s+1)^2} \frac{\rho^s}{\xi^2} + \cdots  \, ,
\label{eq:vfvm}\end{align}
for  $\rho(\xi) \equiv [-a (s+1) \xi]_\Gamma^{1/(s+1)}$,
where the subscript $\Gamma$ indicates that objects inside the brackets are complex and care should be taken with the corresponding phase upon evaluation. The leading term in $V_f(\xi)$ is the $1/\xi^2$ pole such that the near-origin solution a Bessel problem; the next-to-leading term incorporates both leading corrections to the tortoise coordinate and those arising from the metric function. In $V_m(\xi)$ we include only the first correction incorporating angular momentum.

The near-origin master equation generically goes like
\be 
\frac{\dn^{2}R}{\dn z^{2}} + \left[1 + \frac{1/4 - \nu^2}{z^2} + \lambda z^{q-2} \right]R = 0\;,
\ee
where $\lambda$ is a complex constant which we take to have small modulus, and $0 < q < 1$ is real. The leading order corrections from the metric and the angular momentum terms can both have this form.  We now solve for the leading correction to the QNM quantization condition~(\ref{eq:qnm_quant_leading}).

Identify $\mathcal{H}_0 = \partial_z^2 + 1 + (1/4-\nu^2)/z^2$ and decompose $R = R^{(0)} + \lambda R^{(1)}$.  Collecting powers of $\lambda$ yields
\be 
\mathcal{H}_0 R^{(0)} = 0 \, , \qquad \mathcal{H}_0 R^{(1)} = -z^{q-2} R^{(0)} \, .
\ee
Evidently, the zeroth-order solution sources the correction. We solve this problem using variation of parameters.  Let $R_\pm(z)$ be the two independent solutions to $\mathcal{H}_0 R^{(0)} = 0$. We have $R_\pm (z) = \sqrt{2 \pi z} \, J_{\pm \nu}(z)$, with Wronskian ${\cal W}[R_+, R_-] = -4 \sin (\pi\nu)$. We then write $R^{(1)} = c_1(z) R_+ + c_2(z) R_-$ and impose $c_1' R_+ + c_2' R_- = 0$ and find that the equation reduces to
\be \label{eq:subleading_corr_eq}
c_1' R_+' + c_2' R_-' = -  z^{q-2} R^{(0)} \, .
\ee
To solve, it is useful to introduce integrals
\be 
H_{\sigma \tau} (z; q) = \int_{+\infty}^z {\rm d}t\, t^{q-1} J_{\sigma \nu}(t) J_{\tau \nu}(t) 
\ee
where $\sigma, \tau \in \{\pm 1\}$. Convergence at the lower limit occurs when $q < 1$, as holds in our case. 
Solving \eqref{eq:subleading_corr_eq} we obtain 
\begin{align}
&R(z) =\left[C_1  - \frac{\lambda \pi}{2 \sin (\pi \nu)} \left(C_1 H_{+-} + C_2 H_{--} \right)\right] R_+
\nonumber 
\\
&+ \left[C_2 + \frac{\lambda \pi}{2 \sin (\pi \nu)} \left(C_1 H_{++} + C_2 H_{+-}\right) \right]R_-\;. \label{eq:Rsol_subleading_pert_gen}
\end{align} 
Here we have chosen integration constants $c_{1,2}(\infty) = 0$ to ensure the large-$z$ limit of the solution agrees with the generic solution near the origin. As a consequence, matching to the large-$r$ solution is unaltered, leading to
\be \label{eq:infin-orgin-match}
\frac{C_2}{C_1} = - \frac{\sin \left(\omega \xi_0 - \frac{\pi \nu}{2} \right)}{\sin \left(\omega \xi_0 + \frac{\pi \nu}{2} \right)} \, .
\ee


We again match solutions at infinity and the horizon by following a Stokes ray from infinity to $r = 0$, performing a counterclockwise rotation in the complex $z$-plane by $\pi$, and then following a Stokes ray to the horizon. For the leading-order solution, this amounts to analytical continuation of the Bessel functions with the only subtlety arising from the rotation near $r = 0$, which gives $R_\pm \to e^{2 i \alpha_\pm} R_\pm$. The perturbed solution is more subtle because we must track the integrals along the contour so that we can identify the accumulated phase difference.

Leaving details to the supplemental material (see also \cite{Coviello:2026hdc}), we find that near the horizon and in a large-$z$ expansion yields plane wave asymptotics,
\be 
R^{(II)}(z) \sim B^{(II)}_+ e^{i \omega \xi} + B^{(II)}_- e^{-i \omega \xi}\;,
\ee
with coefficients
\begin{align}
B^{(II)}_- &= C_1 e^{i \alpha_+} + C_2 e^{i \alpha_-} \,,
\\
B^{(II)}_+ &=  C_1 e^{3 i \alpha_+} + C_2 e^{3 i \alpha_-} + \pi \lambda \Delta_{+-} \left(C_1 e^{i \alpha_+} + C_2 e^{i \alpha_-} \right)\nonumber
\end{align}
where $\Delta_{\sigma \tau}(q) \equiv \int_\Gamma {\rm d}t \, t^{q-1} J_{\sigma \nu}(t) J_{\tau \nu}(t)$ for contour $\Gamma$. The QNM boundary conditions (\ref{eq:asympsolns}) require $B_+^{(II)} = 0$, together with the matching~\eqref{eq:infin-orgin-match}. 

To solve for the corrected QNMs, define $X = \omega \xi_0$ and
\be 
\eta(X) \equiv - \frac{\sin \left(X - \pi \nu/2 \right)}{\sin \left(X + \pi \nu/2 \right)}  = \frac{C_2}{C_1} \, .
\ee
The horizon ingoing boundary condition becomes
\be 
\eta(X) + e^{3 i \pi \nu} - i \pi \lambda \Delta_{++} \left[ e^{i \pi \nu} + \eta(X)\right] = 0 + O(\lambda^2).
\ee
Linearizing $X = X_0 + \delta X$ we obtain the perturbative correction to the leading QNM quantization (\ref{eq:qnm_quant_leading})
\be 
\hspace{-3mm}\delta \omega\xi_0  = \frac{i \pi^2 e^{\frac{i \pi q}{2}} \Gamma(1-q) \lambda}{2^{2-q} \cos(\pi\nu) \Gamma^2(1-\frac{q}{2}) \Gamma(1+\nu -\frac{q}{2}) \Gamma(1-\nu-\frac{q}{2})}.
\ee
Note that $\lambda$ depends on $\omega$, however, in the leading-order correction, we substitute $\omega \to  n \pi/\xi_0$.

Let us specialize to metric and angular momentum corrections by substituting in appropriate values of $\lambda$ and $q$ (cf. supplemental material). For the metric correction,
\be 
\delta \omega_f = \frac{\pi c p (s-p) \Delta_{++} n^{-\frac{p}{(s+1)}}}{4 (s+1)^2 (s+p+1) \xi_0 (1 + e^{2 i \pi \nu})}  \left[\chi\right]_\Gamma^{\frac{p}{(s+1)}},
\ee
for $\chi\equiv-\frac{a(s+1) \xi_0}{\pi}$, while angular momentum yields
\be \label{eq:ang_mom_correction_gen}
\delta \omega_m = \frac{i m^2 \pi^2 e^{i \pi \nu} n^{-2 \nu }}{2^{2-2\nu} a (s+1)^2 \xi_0 \cos(\pi \nu) \Gamma^2(1-\nu)} \left[\chi\right]_\Gamma^{2 \nu},
\ee
with $\nu = \frac{s}{2 (s+1)}.$

\noindent \emph{Extracting singularity data.} 
The advantage of the subleading corrections is, though their coefficients  depend on the Stokes topology, their scaling with $n$ provides a way to identify the singularity scaling independent of the Stokes topology. 
To see this, consider corrections proportional to $m^2$ (\ref{eq:ang_mom_correction_gen}), 
$\delta \omega_{m,n} \propto m^2 n^{-2 \nu},$
where the proportionality constant depends on the behavior of the Stokes lines. (We calculated this for a Schwarzschild-AdS-like Stokes topology.) Given a QNM spectrum, for $m_1^2 \neq m_2^2$, we form the combination
\be \label{eq:Delta_two_mode}
\Delta_n \equiv \omega_{m_1,n} - \omega_{m_2, n} \propto (m_1^2-m_2^2) \,  n^{-2\nu}  + \cdots \, .
\ee
Fitting the resulting data one can extract $\nu$, and hence the scaling of the metric function near the singularity.

In particular, we can apply our reasoning to extract the Kasner exponents from the large overtone QNM data. To do so for interesting geometries, we first generalize our metric (\ref{eq:sphersymmmet}) to include a second blackening factor
\be 
{\rm d}s^2 = -f(r) {\rm d}t^2 + g^{-1}(r){\rm d}r^2 + r^2 {\rm d}\phi^2 \, .
\ee
Assuming the spacetime is asymptotically AdS$_{3}$, and for a near-singularity metric where 
\be 
f(r)  \sim  -a r^{-s_1}  \, , \quad g(r) \sim - b r^{-s_2}\,,  \qquad s_1, s_2 > 0 \, ,
\ee
our scaling argument for large overtone QNMs goes through straightforwardly. We find (see supplemental)
\be 
\delta \omega_{m,n} \propto m^2 n^{-q_m} \, , \quad \text{where} \quad q_m = \frac{2 s_2}{s_1+s_2+2} \, .
\ee
For $0 < q_m < 1$, this is the leading angular momentum correction in the perturbative regime. For $s_1 = s_2 = s$ we recover scaling $n^{-\frac{s}{(s+1)}}$. Note that with two blackening factors the angular momentum corrections cannot individually disentangle the scaling exponents. 

Take $a, b > 0$ such that the curvature singularity is spacelike. On approach to the singularity, $r$ is timelike, and we can cast the local metric into a Kasner form (\ref{eq:Kasgeomgen}) via the local proper time coordinate change, $\tau \propto r^{(s_2 + 2)/2}$, resulting in the near-singularity metric
\be 
{\rm d} s^2 = - {\rm d}\tau^2 + \tau^{2p_t} {\rm d}t^2 + \tau^{2 p_\phi} {\rm d}\phi^2 \, ,
\label{eq:kasform}\ee
with Kasner exponents
\be 
p_t = -\frac{s_1}{s_2+2} \,, \qquad p_\phi = \frac{2}{s_2+2} \, .
\ee
When there is only one blackening factor and $s_{1}=s_{2}$, the Kasner exponents can be directly read from the asymptotic QNMs. With two blackening factors only
\be 
q_m = \frac{2(1-p_\phi)}{1 - p_t} \, ,
\ee
can be read off.
(This result was independently uncovered in \cite{Hartnoll:2026vhu} as this Letter was in preparation.)
To obtain both Kasner exponents more is needed, e.g., in certain theories of gravity the sum of Kasner exponents has a fixed, solution-independent value~\cite{Bueno:2024qhh}. Thus, with knowledge of the theory, the  exponents can be disentangled.  


\noindent \textbf{Probing a quantum black hole singularity.} Let us now apply our generic recipe for extracting singularity data to an exact, quantum black hole in AdS$_{3}$, namely, the quantum BTZ (qBTZ) black hole \cite{Emparan:2020znc}. This solution arises when an end-of-the-world brane \cite{Karch:2000ct} intersects the horizon of an accelerating AdS$_{4}$ black hole \cite{Emparan:1999wa,Emparan:1999fd}.

\noindent \emph{Geometry of quantum BTZ.} The metric function of the static qBTZ of mass $M$ is \cite{Emparan:2020znc} (cf. \cite{Panella:2024sor} for a review)
\beq
\begin{split}
&f(r)=\frac{r^2}{\ell_3^2}-8\mathcal{G}_3M-\frac{\ell F(M)}{r}\,.
\end{split}
\label{eq:qBTZmain}\eeq
Here, 
$\ell>0$ is an ultraviolet cutoff length scale, 
  $\mathcal{G}_3=G_{3}/\sqrt{1+(\ell/\ell_{3})^{2}}$ is the `renormalized' Newton's constant, and $F(M)$ is a positive function of the mass.

  \begin{figure}[t!]
    \centering
    \includegraphics[width=.47\textwidth]
        {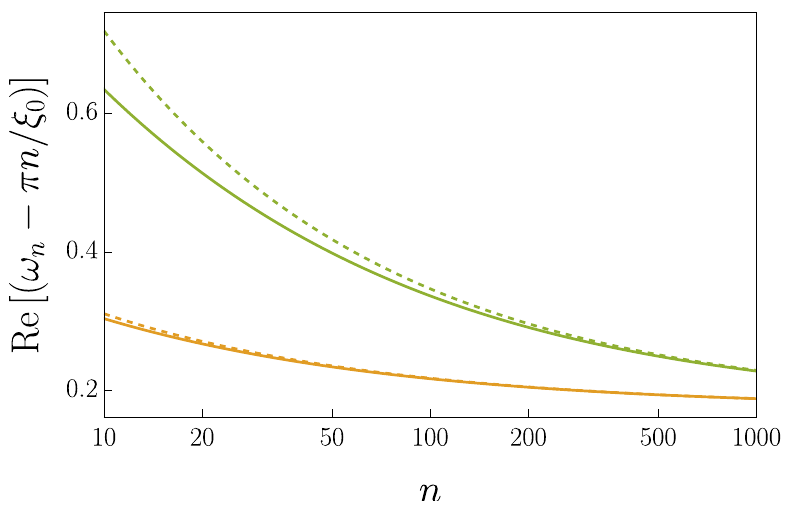}
        \includegraphics[width=.47\textwidth]
        {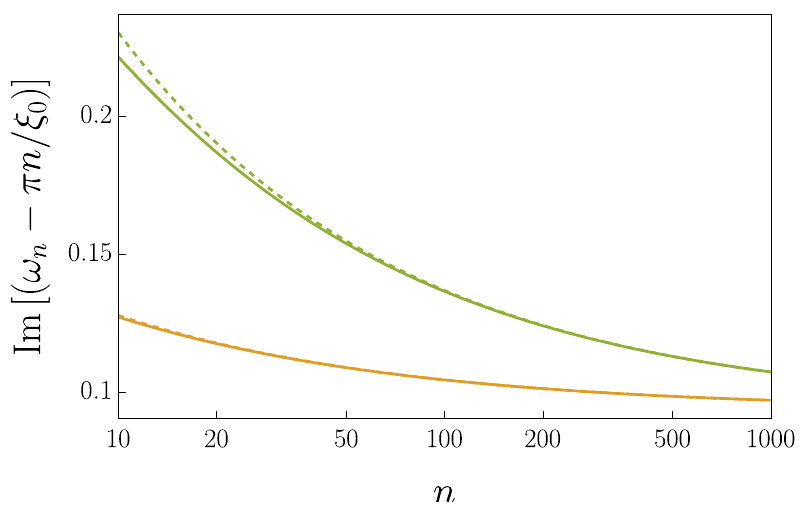}
    \caption{Comparison of numerically computed highly damped QNM
frequencies (solid lines) with the analytically predicted corrections (dashed lines) for $\ell=10$. To make the subleading behavior apparent, we subtract the leading contribution. Bottom (top) pair of curves has angular number $m=1$ ($m=2$).
\label{fig:compare_numerical_analytical_neutral}}
\end{figure} 

  According to a being on the brane, metric (\ref{eq:qBTZmain}) is understood to be an exact solution to the full semiclassical, higher-curvature (schematically denoted by $\mathcal{R}$) gravity,
  \beq \hspace{-3mm}I=\hspace{-1mm}\frac{1}{16\pi G_{3}}\hspace{-1mm}\int \hspace{-1mm} d^{3}x\sqrt{-g}\left(\frac{2}{\ell_{3}^{2}}+R+\ell^{2}\mathcal{R}^{2}+\cdots\right)\hspace{-1mm}+\hspace{-1mm}I_{\text{CFT}}.\label{eq:branetheory}\eeq
In this view, parameter $\ell$ controls the strength of backreaction due to the $\text{CFT}_{3}$ of central charge $c_{3}\gg1$. Small backreaction implies $\ell/\ell_{3}\ll1$. In this regime, gravity becomes weak on the brane, and when $\ell\to0$, we recover the classical BTZ black hole \cite{Banados:1992wn,Banados:1992gq}. Notably, the solution (\ref{eq:qBTZmain}) is exact for \emph{any} $\ell$, even when quantum effects are large $\ell\gg1$. 
  As $\ell$ becomes large the higher-derivative terms become important; in a resummation, the theory (\ref{eq:branetheory}) is equivalent to AdS$_{4}$-Einstein gravity.

\noindent \emph{Kasner eons: classical-to-quantum black hole transition.} From our recipe, we can compute the asymptotic QNM spectrum for a scalar probe of the  neutral qBTZ spacetime. The qBTZ metric (\ref{eq:qBTZmain}) fits the ansatz (\ref{eq:metric_f_ansatz}) for  $s = 1$ and $p = s$. Consequently, the leading correction to the QNM spectra coming from the metric vanishes. Thence, the asymptotic QNM frequency with subleading (momentum) corrections follows from (\ref{eq:ang_mom_correction_gen})
\be 
\omega_{m,n}\xi_{0} = n \pi + \frac{i}{4} \ln 2 + \frac{e^{i \pi/4} \Gamma^2(1/4)}{8 \sqrt{2 \pi }} m^2 \sqrt{\frac{\xi_0}{n\ell F(M)}}.
\ee
In Fig.~\ref{fig:compare_numerical_analytical_neutral} we compare our analytical result to the numerical QNM spectrum for a particular value of $\ell$ (other values appear in \cite{Coviello:2026hdc}). Agreement  improves as the overtone $n$ increases; convergence to the asymptotic approximation is faster for large values of $\ell$ and small values of $m$.

For sufficiently small $\ell/\ell_3$, the interior of the qBTZ black hole approximates the classical BTZ interior over the radial range $\ell F  \ll r \ll  \ell_3$. A separation of scales develops between the two endpoints when $\ell/\ell_3 \ll 1$.  In this regime, the interior geometry has the Kasner
form (\ref{eq:kasform}) with exponents $p^{\text{BTZ}}_t = 0$ and $p^{\text{BTZ}}_\phi = 1$. These
exponents describe the approach to the horizon of a Milne patch,
rather than a curvature singularity (indeed, the BTZ solution has no curvature singularity). Deeper in the interior, however, quantum corrections dominate and the geometry approaches a Kasner singularity with exponents $p^{\text{qBTZ}}_t = -1/3$ and $p^{\text{qBTZ}}_\phi = 2/3$. This is the qBTZ dominated regime, which will always be reached at sufficiently small $r$. 
The crossover between these  regimes signals a transition between Kasner eons~\cite{Bueno:2024fzg}. 

\begin{figure}[h!]
\centerline{\includegraphics[width=.42\textwidth]{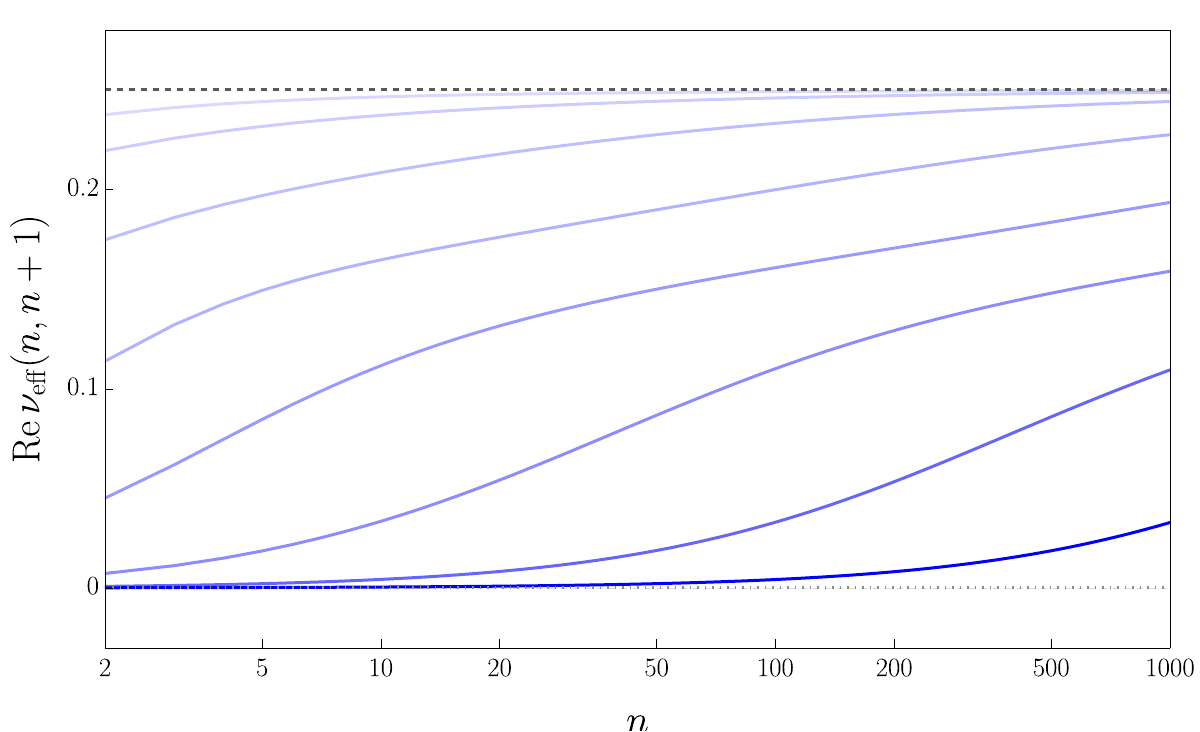}}
    \caption{{\label{fig:eon}} Kasner eon transition of classical to quantum BTZ black hole. Curves correspond to values of $\ell$ from $10^{-4}$
(lowest blue curve) to $10^{3}$ (highest blue curve), increasing by
factors of ten as the opacity decreases. Horizontal lines indicate analytically derived scaling exponent $\nu$ for BTZ (lower, dotted)
and for qBTZ in the large-overtone limit (upper, dashed). For small
$\ell$, a pronounced BTZ ``classical eon'' emerges, with the effective scaling
exponent matching the BTZ value of $\nu = 0$. As the overtone number increases
at fixed $\ell$, the system transitions from a BTZ-dominated to a
qBTZ-dominated ``quantum eon'', corresponding to a transition between
Kasner eons in the quantum black hole. For larger $\ell$, the BTZ
regime is absent, and quantum corrections dominate from the outset. For very small values of $\ell$, convergence to the qBTZ prediction would take hundreds of thousands of modes.}
\end{figure}

The transition between Kasner eons should be reflected in the highly damped QNM spectrum. To this end, we work with an effective scaling exponent 
\be 
\nu_{\rm eff}(n, n+1) = - \frac{\log \left( |\Delta_{n+1}|/|\Delta_n| \right)}{2 \log \left((n+1)/n\right)}.
\ee
for $\Delta_n$ \eqref{eq:Delta_two_mode}. When the asymptotic approximation holds, we have $\nu_{\rm eff} \to \nu$ where $\nu$ is the genuine scaling exponent associated with the singularity. An exact BTZ interior has $\nu = 0$, while for the qBTZ regime we have $\nu = 1/4$.  We display the transition between Kasner eons in Fig.~\ref{fig:eon}.

\noindent \textbf{Discussion.} Here we supplied a robust method showing how to reconstruct data of a black hole singularity, specifically, the scaling exponent, from the asymptotic QNM spectra. We then applied our method and explicitly uncovered a transition between Kasner eons inside a quantum black hole. Our work thus precisely realizes an instance of emergent physics as one probes deeper into a black hole interior, and that, in principle, such information could be seen by an external observer. 

Our work opens up new avenues to explore. This includes extracting singularity data for other types of black holes, such as static quantum black holes with flat or de Sitter asymptotics \cite{Emparan:2022ijy,Climent:2024nuj,Climent:2024wol,Feng:2024uia,Bhattacharya:2025tdn}. Our method readily extends to higher-dimensions (this was achieved for AdS$_{5}$ black holes \cite{Xiao:2026pir}), or the inclusion of matter for which one could track other novel interior transitions \cite{Frenkel:2020ysx,Caceres:2024edr,Caceres:2026mug}. It would also be worth connecting to other explorations into the near-singularity dynamics of AdS black holes and their holographic interpretation, e.g., \cite{Hartnoll:2020rwq,Hartnoll:2020fhc,Sword:2021pfm,Sword:2022oyg,
Caceres:2022smh,Hartnoll:2022snh,Caceres:2022hei,Caceres:2023zhl,Caceres:2023zft,DeClerck:2023fax,Blacker:2023ezy,Carballo:2024hem}, or how asymptotic QNMs relate to holographic OPE data of appropriate two-point functions \cite{Fidkowski:2003nf,Festuccia:2005pi,Ceplak:2024bja,Afkhami-Jeddi:2025wra,Jia:2026ryl,Ceplak:2026ard}.

\noindent
{\bf Note added.} In the final stages of preparing this manuscript, Refs.~\cite{Hartnoll:2026vhu, Xiao:2026pir} appeared, which overlap with our work on extracting singularity data from the difference in frequencies at different angular momentum quantum number.  

\noindent \emph{Acknowledgements.}
We are grateful to  Ruth Gregory and Juan Pedraza for insightful feedback.
 CC and AG are supported by King’s College London through an NMES funded studentship. KS is supported by a scholarship from King's-China Scholarship Council. AS is funded by the Royal Society under the grant “Concrete Calculables in Quantum de Sitter” and  further supported by the STFC consolidated grant ST/X000753/1.

\bibliographystyle{apsrev4-2}
\bibliography{qBHrefs}

\newpage

\onecolumngrid

\section{Supplemental material}

\subsection{Leading asymptotic QNMs}

\noindent We follow the strategy outlined in the main text to solve for the asymptotic QNMs and their subleading corrections. 

\noindent \emph{Large-$r$ solution.} At large radius, the potential behaves as 
\be 
V_{(\infty)}  = \frac{3 r^2}{4 \ell_3^4} + O(1) \, .
\ee
In the same limit, the tortoise coordinate evaluates to
\be 
\xi - \xi_0 = - \frac{\ell_3^2}{r} + O(r^{-3}) \, .
\ee
Now let $z = \omega (\xi_0 - \xi)$. To leading order, the general solution can be written as
\be 
R_{(\infty)}(z) = A_1 \sqrt{2 \pi  z} \, J_1(z) + A_2 \sqrt{2 \pi z} \, Y_1(z) \, .
\ee

The requirement of normalizable boundary conditions at $r \to \infty$ requires us to set $A_2 = 0$. Expanding the resulting solution at large argument allows us to connect the result to the plane wave basis, as stated in the main text.

\noindent \emph{Near horizon solution.} At the horizon 
the tortoise coordinate behaves logarithmically,
\be
\xi = \frac{1}{f'(r_h)} \log (r - r_h) + O(r-r_h) \, .
\ee
 and the potential vanishes linearly,
\be 
V_h = f'(r_h) \left[\frac{m^2}{r_h^2} + \frac{f'(r_h)}{2 r_h} \right] (r-r_h) + O \left((r-r_h)^2 \right) \, .
\ee
Hence, the local solution exhibits plane wave behavior, stated in the main text.
\be 
R_{(h)}(\xi) \sim B_1 e^{i \omega \xi} + B_2 e^{-i \omega \xi } \, .
\ee

\noindent \emph{Near origin solution.} Near the origin,  
the leading-order differential equation approximates to
\be 
\frac{\dn^{2}R}{\dn z^{2}} + \left(1 + \frac{\frac{1}{4}-\nu^2}{z^2} \right) R = 0 \,, \qquad \text{where} \qquad \nu = \frac{s}{2 (s+1)} \, , 
\ee
with general solution
\be \label{eq:R0_gen_sol}
R_{(0)}(z) = C_1 \sqrt{2 \pi  z} J_\nu (z) + C_2 \sqrt{2 \pi  z} J_{-\nu }(z) \, .
\ee
Since $s > 0$, the index of the Bessel function will never be an integer, and $\sqrt{2 \pi z} J_{\pm \nu}(z)$ provides an adequate basis. Again, at large argument, the solution can be expanded in a plane wave basis.

\noindent \emph{Stokes rays and topology.} Thus far we have constructed local solutions near infinity, near the horizon, and near the origin. To obtain the QNM quantization condition, we match the plane-wave solutions and impose boundary conditions. Generally, however, such a matching is difficult because the solutions mix growing and decaying exponential terms. To circumvent this difficulty, we match along the Stokes line ${\rm Im}(\omega \xi) = 0\;.$ First, near infinity, the Stokes ray is ${\rm Im}(\omega \xi_0) = 0$. 
As a matter of convention, on the Stokes ray connected to infinity, we take ${\rm Re}(\omega \xi) > 0$. Then, ${\rm Arg}(\omega) = - {\rm Arg}(\xi_0)$.
Given the relation between $\xi$ and $r$, there is a single Stokes ray extending to the AdS boundary. 

Next, near a `horizon' (a real or complex root to the blackening factor), $r=r_{h}$, we have $\xi \propto {\rm log}(r-r_h) $. Let $r - r_h = \rho e^{i \vartheta}$. Then a Stokes line near the horizon satisfies
\be 
\vartheta = - \frac{{\rm Im}(\omega/f'(r_h))}{{\rm Re}(\omega/f'(r_h))} \, \log \rho + \, \text{const}. \, ,
\ee
thereby spiraling into the horizon in the complex-$r$ plane.

Finally, near the origin the tortoise coordinate admits a local expansion. Solving for the Stokes lines gives,
\be \label{eq:stokes_sep_gen}
{\rm Arg}(r) = \frac{\pi k}{s+1} + \frac{{\rm Arg(a)}}{s+1} - \frac{{\rm Arg}(\omega)}{s+1} \, , \qquad k = 0, 1, 2, \dots \, .
\ee
A number of Stokes rays then emerge from $r = 0$ in the complex plane, separated by $\pi/(s+1)$. For integer $s$, there are $2(s+1)$ total rays emanating from the origin, while for non-integer $s$ there will be a branch cut at the origin. 

So far we have only described the \textit{local} geometry of the Stokes lines. Let us now understand how the Stokes lines connect. Doing so requires global information about the metric function. Let $\sigma = \omega \xi \in \mathbb{R}$ be the coordinate along the Stokes ray. Given the metric function $f(r)$, the Stokes rays can be obtained by solving
\be 
\sigma \equiv \omega \xi = \omega \int^r \frac{{\rm d} \bar{r}}{f(\bar{r})} \qquad \Longrightarrow \qquad \frac{{\rm d} r}{{\rm d}\sigma} = \frac{f(r)}{\omega} \, .
\ee
Only the complex phase of $\omega$ is needed to solve this differential equation. Thus, without loss of generality, we determine the Stokes lines from solving 
\be 
\frac{{\rm d} r}{{\rm d}\sigma} = \frac{f(r)}{e^{-i {\rm Arg}(\xi_0)}} \, . 
\ee
In summary, given an explicit choice for the metric function $f(r)$, the above procedure allows for the construction of the Stokes lines and connected contour. To do so for a more general metric function, we \textit{assume} the Stokes topology is Schwarzschild-AdS-like. Specifically, the relevant contour connecting infinity to the horizon has the structure: follow the Stokes line from infinity to $r = 0$, rotate counterclockwise by $\pi/(s+1)$ in the complex radius plane, and then follow the corresponding Stokes line to the horizon.

\noindent \emph{Solution matching and QNM quantization.} Given the assumption for the contour described above, we need to transform the solution near the origin~\eqref{eq:R0_gen_sol} by a $\pi/(s+1)$ rotation in the complex-$r$ plane. This amounts to transforming $z \to z e^{i \pi}$, such that
\be 
R_{(0)}^{(II)}(z) = C_1 e^{i 2 \alpha_+} \sqrt{2 \pi z} J_\nu (z) + C_2 e^{i 2 \alpha_-} \sqrt{2 \pi z} J_{-\nu} (z) \, .
\ee
We then expand the solution at large argument, and express in the plane wave basis. 
 Expanding at large argument gives
\be \label{eq:gen_hor_R_cont}
R_{(0)}^{(II)}(z) \sim \left(C_1 e^{3 i \alpha_+} + C_2 e^{3 i \alpha_-} \right) e^{i \omega \xi} + \left( C_1 e^{i \alpha_+} + C_2 e^{ i \alpha_-} \right) e^{-i \omega \xi} \, .
\ee
Matching the solutions on the infinity-to-origin portion of the contour gives the match (\ref{eq:infin-orgin-match}). Using this result in~\eqref{eq:gen_hor_R_cont} and setting the outgoing contribution to vanish yields the asymptotic QNM quantization condition.

\subsection{Subleading corrections to asymptotic QNMs}

\noindent Here we provide details for computing the subleading corrections to the asymptotic QNM spectrum. 

\noindent \emph{Subleading corrections to the effective potential.} Given the near-origin behavior of~\eqref{eq:metric_f_ansatz}, we can first identify the correction to the tortoise coordinate,
\be 
\xi = - \frac{r^{s+1}}{a(s+1)} \left[1 - \frac{c(s+1)}{s+p+1} r^p + \cdots  \right] \, .
\ee
For notational convenience, let us define
\be 
\rho(\xi) \equiv [-a (s+1) \xi]_\Gamma^{1/(s+1)} \, ,
\ee
with the branch chosen such that $\rho \to r$ on the origin-infinity Stokes ray. The subscript $\Gamma$ serves as a reminder that, because the objects inside the brackets are complex, one must take care with the corresponding phase in evaluating these quantities. Inverting this expression to obtain $r(\rho)$ gives
\be 
r = \rho \left[1 + \frac{c}{s+p+1} \rho^p + \cdots \right] \, .
\ee

Recall for a massless scalar in three-dimensions, the effective potential has the form (\ref{eq:brane-potential-general}). 
We are interested in seeing how the leading corrections to the metric and the angular momentum $m$ contributions modify the near-origin form of the potential. Expanding the potential near $r = 0$ and using~\eqref{eq:metric_f_ansatz} we write $V = V_f + V_m$ with
\begin{align}
    V_f(r) &= a^2 r^{-2(s+1)} \left[-\frac{1+2s}{4} + \frac{c(p-2s-1)}{2} r^p + \cdots \right] \,, 
    \\
    V_m(r) &= -a m^2 r^{-s - 2} \left[ 1 + c r^p + \cdots \right] \, .
\end{align}
After substituting for the tortoise coordinate (and using $\rho$ from above) we obtain (\ref{eq:vfvm}).
 In practice, the leading metric and angular-momentum corrections need not be consecutive terms in the asymptotic expansion of the potential --- additional contributions may occur at intermediate orders. Our aim here, however, is to isolate the leading correction associated with each of these two sources. 

 Thus, near the origin the master equation is (\ref{eq:schroprob1Dz})
\be 
\frac{\dn^{2} R}{\dn z^{2}} + \left[1 + \frac{\frac{1}{4} -\nu^2}{z^2} + \lambda_f z^{q_f - 2} + \cdots + \lambda_m z^{q_m-2} + \cdots \right] R  = 0.
\ee
Here $q_f \equiv \frac{p}{s+1}$ and
\be \label{eq:q_lambda_metric}
 \lambda_f \equiv \frac{c p (s-p)}{2(s+1)^2(s+p+1)} \left[-\frac{a(s+1)}{\omega}\right]_\Gamma^{p/(s+1)} \, ,
\ee
controls the leading correction due to the metric, while $q_m \equiv \frac{s}{s+1} = 2 \nu$ and
\be \label{eq:q_lambda_angmom}
 \lambda_m = \frac{m^2}{a (s+1)^2}\left[-\frac{a(s+1)}{\omega}\right]_\Gamma^{s/(s+1)} \, ,
\ee
controls the leading correction due to momentum $m$.

\noindent \emph{Modified solution matching.} We can solve the (near-origin) master field $R$ radial equation as described in the main text, leading to (\ref{eq:subleading_corr_eq}). We then match the solutions at infinity and the horizon by following a Stokes ray from infinity to $r = 0$, performing a counterclockwise rotation in the complex $z$-plane by $\pi$, and then following a Stokes ray to the horizon.  The perturbed solution is more subtle because we must track the integrals along the contour so that we can identify the accumulated phase difference.

To be clear, denote the specified contour $\Gamma$ and define
\be \label{eq:Delta_gen_contourap}
\Delta_{\sigma \tau}(q) = \int_\Gamma {\rm d}t \, t^{q-1} J_{\sigma \nu}(t) J_{\tau \nu}(t) \, . 
\ee
Evaluating this integral yields \cite{Coviello:2026hdc}
\be 
\Delta_{\sigma \tau}(q) = \left(e^{i \pi [q + (\sigma + \tau)\nu]} - 1\right) I_{\sigma \tau} \, , \quad I_{\sigma \tau} = \int_0^\infty {\rm d}t \, t^{q-1} J_{\sigma \nu}(t) J_{\tau \nu}(t) \, . 
\ee 
where $I_{\sigma \tau}$ is a Weber-Schafheitlin integral~\cite{spWeber} 
\be \label{eq:Weber-Schafheitlin}
\int_0^\infty {\rm d}t\,t^{q-1}J_\alpha(t)J_\beta(t)
=\frac{2^{q-1}\Gamma(1-q)\Gamma((\alpha+\beta+q)/2)}
{\Gamma(1+(\beta-\alpha-q)/2)\Gamma(1+(\alpha+\beta-q)/2)
\Gamma(1+(\alpha-\beta-q)/2)} \, ,
\ee
where convergence requires $2\nu < q < 1$. Extensions of the integrals beyond this domain can be achieved via analytic continuation~\cite{Coviello:2026hdc}. 
Presently,
\be 
I_{++} = \frac{2^{q-1} \Gamma (1-q) \Gamma(\nu +q/2)}{\Gamma^2(1-q/2) \Gamma(1+ \nu - q/2)} \, ,
\ee
and
\be 
\frac{I_{-+}}{I_{++}}=\frac{I_{+-}}{I_{++}} = \frac{\sin \left[\pi (q/2 +\nu) \right]}{\sin (\pi q/2)} \, , \qquad \frac{I_{--}}{I_{++}} = \frac{\sin \left[\pi(q/2 + \nu) \right]}{\sin \left[\pi (q/2 - \nu) \right]} \, .
\ee
Here we have used the identity $\Gamma(x) \Gamma(1-x) = \pi/ \sin (\pi x)$ in simplifying the results. Putting this together we have
\be \label{eq:Delta_vals_horizon}
\Delta_{+-} = e^{-i\pi \nu} \Delta_{++} \,, \quad \Delta_{--} = e^{-2 i \pi \nu} \Delta_{++} \,, \quad \Delta_{++} = 2 i e^{i \pi(q/2+\nu)} \sin \left[\pi(q/2 + \nu)\right] I_{++} \, .
\ee
In the case $q = 2 \nu$, the only potentially problematic term is $\Delta_{--}$, which should then be understood as a limit. Specifically, the part of the contour where the solution is rotated from the infinity-origin curve to the origin-horizon curve makes a non-vanishing contribution and we have
\be 
\Delta_{--}(2\nu) = \frac{i \pi 2^{2\nu}}{\Gamma^2(1-\nu)} \, .
\ee
The behavior of $R(z)$ near the horizon is then given by substituting $H_{\sigma \tau} \to \Delta_{\sigma \tau}$ in solution~\eqref{eq:Rsol_subleading_pert_gen} using the results of ~\eqref{eq:Delta_vals_horizon}. We then find in the large-$z$ expansion plane wave asymptotics.

\end{document}